\documentclass[11pt]{article}
\input epsf
\makeatletter

\def\newpic#1{}

\voffset= - 0.2in
\def\hybrid{\topmargin 0pt      \oddsidemargin 0pt
             \headheight 0pt \headsep 0pt

             \textwidth 6.25in       
             \textheight 9.5in       
             \marginparwidth 0.0in
             \parskip 5pt plus 1pt   \jot = 1.5ex}
\catcode`\@=11
\def\marginnote#1{}

\newcount\hour
\newcount\minute
\newtoks\amorpm
\hour=\time\divide\hour by60
\minute=\time{\multiply\hour by60 \global\advance\minute by-\hour}
\edef\standardtime{{\ifnum\hour<12 \global\amorpm={am}%
             \else\global\amorpm={pm}\advance\hour by-12 \fi
             \ifnum\hour=0 \hour=12 \fi
             \number\hour:\ifnum\minute<10 0\fi\number\minute\the\amorpm}}
\edef\militarytime{\number\hour:\ifnum\minute<10 0\fi\number\minute}

\def\draftlabel#1{{\@bsphack\if@filesw {\let\thepage\relax
        \xdef\@gtempa{\write\@auxout{\string
           \newlabel{#1}{{\@currentlabel}{\thepage}}}}}\@gtempa
        \if@nobreak \ifvmode\nobreak\fi\fi\fi\@esphack}
             \gdef\@eqnlabel{#1}}
\def\@eqnlabel{}
\def\@vacuum{}
\def\draftmarginnote#1{\marginpar{\raggedright\scriptsize\tt#1}}

\def\draftlabel#1{{\@bsphack\if@filesw {\let\thepage\relax
        \xdef\@gtempa{\write\@auxout{\string
           \newlabel{#1}{{\@currentlabel}{\thepage}}}}}\@gtempa
        \if@nobreak \ifvmode\nobreak\fi\fi\fi\@esphack}
             \gdef\@eqnlabel{#1}}
\def\@eqnlabel{}
\def\@vacuum{}
\def\draftmarginnote#1{\marginpar{\raggedright\scriptsize\tt#1}}

\def\draft{\oddsidemargin -.5truein
             \def\@oddfoot{\sl preliminary draft \hfil
             \rm\thepage\hfil\sl\today\quad\militarytime}
             \let\@evenfoot\@oddfoot \overfullrule 3pt
             \let\label=\draftlabel
             \let\marginnote=\draftmarginnote
        \def\@eqnnum{(\theequation)\rlap{\kern\marginparsep\tt\@eqnlabel}%
\global\let\@eqnlabel\@vacuum}  }

\def\numberbysection{\@addtoreset{equation}{section}
             \def\theequation{\thesection.\arabic{equation}}}

\def\underline#1{\relax\ifmmode\@@underline#1\else
             $\@@underline{\hbox{#1}}$\relax\fi}

\def\titlepage{\@restonecolfalse\if@twocolumn\@restonecoltrue\onecolumn
          \else \newpage \fi \thispagestyle{empty}\c@page\z@
             \def\thefootnote{\fnsymbol{footnote}} }

\def\endtitlepage{\if@restonecol\twocolumn \else  \fi
             \def\thefootnote{\arabic{footnote}}
             \setcounter{footnote}{0}}  
\relax

\makeatletter
\newdimen\normalarrayskip              
\newdimen\minarrayskip                 
\normalarrayskip\baselineskip
\minarrayskip\jot
\newif\ifold             \oldtrue            \def\new{\oldfalse}
\def\arraymode{\ifold\relax\else\displaystyle\fi} 
\def\eqnumphantom{\phantom{(\theequation)}}     
\def\@arrayskip{\ifold\baselineskip\z@\lineskip\z@
         \else
         \baselineskip\minarrayskip\lineskip2\minarrayskip\fi}
\def\@arrayclassz{\ifcase \@lastchclass \@acolampacol \or
\@ampacol \or \or \or \@addamp \or
       \@acolampacol \or \@firstampfalse \@acol \fi
\edef\@preamble{\@preamble
      \ifcase \@chnum
         \hfil$\relax\arraymode\@sharp$\hfil
         \or $\relax\arraymode\@sharp$\hfil
         \or \hfil$\relax\arraymode\@sharp$\fi}}
\def\@array[#1]#2{\setbox\@arstrutbox=\hbox{\vrule
         height\arraystretch \ht\strutbox
         depth\arraystretch \dp\strutbox
         width\z@}\@mkpream{#2}\edef\@preamble{\halign
\noexpand\@halignto
\bgroup \tabskip\z@ \@arstrut \@preamble \tabskip\z@ \cr}%
\let\@startpbox\@@startpbox \let\@endpbox\@@endpbox
      \if #1t\vtop \else \if#1b\vbox \else \vcenter \fi\fi
      \bgroup \let\par\relax
      \let\@sharp##\let\protect\relax
      \@arrayskip\@preamble}
\def\eqnarray{\stepcounter{equation}%
                  \let\@currentlabel=\theequation
                  \global\@eqnswtrue
                  \global\@eqcnt\z@
                  \tabskip\@centering
                  \let\\=\@eqncr
     \halign to \displaywidth\bgroup
        \eqnumphantom\@eqnsel\hskip\@centering
        $\displaystyle \tabskip\z@ {##}$%
        \global\@eqcnt\@ne \hskip 2\arraycolsep
             $\displaystyle\arraymode{##}$\hfil
        \global\@eqcnt\tw@ \hskip 2\arraycolsep
             $\displaystyle\tabskip\z@{##}$\hfil
             \tabskip\@centering
        &{##}\tabskip\z@\cr}
\begingroup\ifx\undefined\newsymbol \else\def\input#1 {\endgroup}\fi
\newfont{\hr}{msbm10}
\newfont{\ams}{msam10}

\def\beq{\begin{equation}}
\def\eeq{\end{equation}}
\def\ba{\beq\new\begin{array}{c}}
\def\ea{\end{array}\eeq}
\def\be{\ba}
\def\ee{\ea}

\def\Re{{\rm Re}}

\def\d{\partial}
\def\p{\partial}

\def\ha{{1\over 2}}
\def\Bf#1{\mbox{\boldmath $#1$}}

\def\bomega{{\Bf\omega}}

\def\bZ{{\bf Z}}

\numberbysection
\hybrid

\begin{document}
\setcounter{footnote}{3}
\begin{titlepage}

\title{On the Dirichlet Boundary Problem and
Hirota Equations
\footnote{Based on the talks given by the authors at NATO ARW on {\sl Hirota
equations}, Elba, Italy, September 2002.}}

\author{
A.~Marshakov \thanks{
Theory Department, P.N.Lebedev Physics Institute
and ITEP,
Moscow, Russia} 
\and A.~Zabrodin
\thanks{Institute of Biochemical Physics
and ITEP,
Moscow, Russia}}

\date{May 2003}
\maketitle
\vspace{-7cm}

\centerline{
\hfill FIAN/TD-10/03}
\centerline{
\hfill ITEP/TH-32/03}

\vspace{7cm}

\begin{abstract}
\noindent
We review the integrable structure of
the Dirichlet boundary problem
in two dimensions. The solution to the Dirichlet boundary problem
for simply-connected case is given
through a quasiclassical tau-function,
which satisfies the Hirota equations of the dispersionless Toda
hierarchy, following from properties of the
Dirichlet Green function. We also outline a possible
generalization to the case of multiply-connected domains related
to the multi-support solutions of matrix models.

\end{abstract}

\vfill

\end{titlepage}
\setcounter{footnote}{0}
\section{Introduction: the Green function and
the Hadamard formula}

Solving the Dirichlet boundary problem \cite{C-H}, one
reconstructs a harmonic function in a
bounded domain from its values on the boundary.
In two dimensions, this is
one of standard problems of complex
analysis having close relations to string theory
and matrix models.
Remarkably, it possesses a hidden integrable
structure \cite{MWZ}.
It turns out that variation of a solution
to the Dirichlet problem under variation
of the domain is described by an infinite
hierarchy of non-linear partial differential equations
known (in the simply-connected case) as dispersionless Toda hierarchy.
It is a particular example of the universal hierarchy of
quasiclassical or Whitham
equations introduced in \cite{KriFun,KriW}.

The quasiclassical tau-function (or its logarithm $F$)
is the main new object associated with a family of domains
in the plane. Any domain in the complex plane with sufficiently
smooth boundary can be parametrized by its
harmonic moments and the $F$-function is a function of the full infinite
set of the moments. The first order
derivatives of $F$ are then moments of the complementary
domain. This gives a formal solution to the inverse potential
problem, considered for a simply-connected case in \cite{M-W-Z,W-Z}.
The second order derivatives
are coefficients of the Taylor expansion of the
Dirichlet Green
function and therefore they solve the
Dirichlet boundary problem. These coefficients are constrained by
infinite number of universal (i.e. domain-independent)
relations which, unified in a generating form, just
constitute the dispersionless Hirota equations. For the third order
derivatives there is a nice ``residue formula''
which allows one to prove \cite{BMRWZ} that $F$ obeys the
Witten-Dijkgraaf-Verlinde-Verlinde (WDVV) equaitons \cite{WDVV}.

Let us remind the formulation of the
Dirichlet problem in planar domains.
Let ${\sf D^c}$ be a domain in the complex plane
bounded by one or several non-intersecting curves.
It will be convenient for us to realize the ${\sf D^c}$
as a complement of another domain,
${\sf D}$ (which in general may have
more than one connected components), and
consider the Dirichlet problem in ${\sf D^c}$.
The problem is to find a harmonic
function $u(z)$ in ${\sf D^c}$, such that it is continuous
up to the boundary $\d {\sf D^c}$ and equals a given function
$u_0(\xi )$ on the boundary, and it can be
uniquely solved in terms of the Dirichlet
Green function $G(z,\xi )$:
\be\label{Dirih}
u(z)=
- \frac{1}{2\pi}\oint_{\d {\sf D^c}}
u_0(\xi )\p_{n} G(z,\xi ) |d\xi |
\ee
where $\p_n$ is the normal derivative on the boundary
with respect to the second variable,
and the normal vector $\vec n$ is directed
inward ${\sf D^c}$, $|d\xi| := dl( \xi)$ is an infinitesimal
element of the length of the boundary $\p {\sf D^c}$.

The Dirichlet Green function
is uniquely determined by the following properties \cite{C-H}:
\begin{itemize}
\item[($G1$)] The function
$G(z,z')$
is symmetric and harmonic everywhere in ${\sf D^c}$
(including $\infty$ if ${\sf D^c} \ni \infty$) in
both arguments except $z=z'$ where
$G(z,z')=\log |z-z'| +\ldots $ as $z\to z'$;
\item[($G2$)] $G(z,z')=0$ if any
of the variables $z$, $z'$ belongs to the boundary.
\end{itemize}
Note that the definition implies that $G(z,z')<0$
inside ${\sf D^c}$. In particular, $\p_n G(z, \xi)$ is
strictly negative for all $\xi \in \p {\sf D^c}$.

If ${\sf D^c}$ is simply-connected (the boundary has only
one component),
the Dirichlet problem is equivalent to finding
a bijective conformal map from
${\sf D^c}$ onto the unit disk
or any other reference domain (where the Green function
is known explicitly) which exists by virtue
of the Riemann mapping theorem.
Let $w(z)$ be such a bijective
conformal map of ${\sf D^c}$ onto the complement to
the unit disk, then
\be\label{Gconf}
G(z, z')=\log \left |
\frac{w(z)-w(z')}{w(z)\overline{w(z')} -1} \right |
\ee
where bar means complex conjugation.
It is this formula which allows one to derive the Hirota
equations for the tau-function of the Dirichlet problem in the most
economic and transparent way \cite{MWZ}.
Indeed, the Green function is shown to
admit a representation
through the logarithm of the tau-function of the form
\be\label{GGG}
G(z,z')=\log \left |\frac{1}{z}-\frac{1}{z'}\right |
+\frac{1}{2}\nabla (z)\nabla (z')F
\ee
where $\nabla (z)$ (see (\ref{D2}) below) is
certain differential operator
with constant coefficients (depending only on the point $z$
as a parameter) in the space of harmonic moments.
Taking into account that $G(z, \infty )=-\log |w(z)|$,
one excludes the Green function from these relations
thus obtaining a closed system of equations for $F$ only.

\begin{figure}[tb]
\epsfysize=5cm
\centerline{\epsfbox{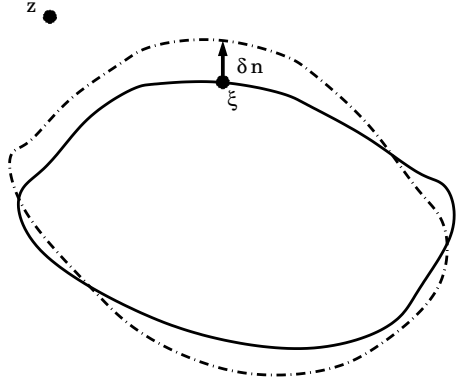}}
\caption{\sl A ``pictorial" derivation of the
Hadamard formula. We consider
a small deformation of the domain, with the new boundary being
depicted by the dashed line. According to ($G2$) the
Dirichlet Green function vanishes $G(z,\xi)=0$
if $\xi$ belongs to the old boundary. Then
the variation $\delta G(z,\xi)$ simply equals to
the new value, i.e. in the leading order
$\delta G(z,\xi) = -\delta n(\xi)\d_n G(z,\xi)$.
Now notice that $\delta G(z, \xi)$ is a {\it harmonic
function} (the logarithmic singularity cancels since it is
the same for both old and new functions)
with the boundary value
$-\delta n(\xi)\d_n G(z,\xi)$.
Applying (\ref{Dirih})
one obtains (\ref{Hadam}).}
\label{fi:hadamar}
\end{figure}

Our main tool to derive (\ref{GGG})
is the Hadamard variational formula \cite{Hadamard} which gives
variation of the Dirichlet Green function under small deformations
of the domain in terms of the Green function itself:
\be\label{Hadam}
\delta G(z, z')=\frac{1}{2\pi}\oint_{\d {\sf D^c}}
\p_{n}G(z, \xi)\p_{n}G(z', \xi)\delta n(\xi)|d\xi |.
\ee
Here $\delta n(\xi)$ is the normal displaycement (with sign)
of the boundary under the deformation,
counted along the normal
vector at the boundary point $\xi$.
It was shown in \cite{MWZ} that this remarkable formula
reflects all
integrable properties of the Dirichlet problem. An
extremely simple ``pictorial" derivation of the Hadamard formula
is
presented in fig.~\ref{fi:hadamar}.
Looking at the figure and applying (\ref{Dirih}), one immediately
gets (\ref{Hadam}).

\section{The Dirichlet problem for simply-connected
domains and dispersionless Hirota equations
\label{ss:simply}}

Let ${\sf D}$ be a connected domain in the complex plane
bounded by a simple analytic curve.
We consider the exterior Dirichlet problem in
${\sf D^c}={\bf C}\setminus {\sf D}$ which is the complement of
${\sf D}$ in the whole (extended) complex plane.
Without loss of generality,
we assume that ${\sf D}$ is compact and
contains the point $z=0$.
Then ${\sf D^c}$ is an unbounded simply-connected domain
containing $\infty$.

\subsection{Harmonic moments and
elementary deformations}

To characterize the shape of the domain ${\sf D^c}$ we
consider its moments with respect to a complete basis
of harmonic functions. The simplest basis is
$\{ z^{-k} \}$, $\{ \bar z^{-k} \}$ ($k\geq 1$)
and the constant function.
Let $t_k$ be the harmonic moments
\be\label{momt}
t_k= \, - \, \frac{1}{\pi k}
\int_{{\sf D^c}}z^{-k} \,d^2 z\,, \,\;\;\;\;\;k=1,2,\ldots
\ee
and ${\bar t}_k$ be the complex conjugated moments.
The Stokes formula
represents them as contour
integrals
$t_{k}=\frac{1}{2\pi i k}\oint_{\d {\sf D}}
z^{-k} \bar z dz$,
providing, in particular, a regularization of possibly divergent
integrals (\ref{momt}). The moment of constant function
is infinite but its variation is always finite
and opposite to the variation of the
complimentary domain ${\sf D}$.
Let $t_0$ be the area (divided by $\pi$)
of ${\sf D}$:
\be\label{t0}
t_0 =
\frac{1}{\pi}\int_{{\sf D}} d^2z
\ee

The harmonic moments of ${\sf D^c}$ are coefficients of
the Taylor expansion of the potential
\be\label{dd1}
\Phi (z) =-\frac{2}{\pi}\int_{{\sf D}} \log |z-z'|d^2z'
\ee
induced by the domain ${\sf D}$ filled by two-dimensional
Coulomb charges with the uniform density $-1$.
Clearly, $\d_z \d_{\bar z}\Phi (z) =-1$ if $z\in {\sf D}$ and vanishes
otherwise, so around the origin (recall that ${\sf D}\ni 0$)
the potential is $-|z|^2$ plus a harmonic function:
\beq\label{deftk}
\Phi (z)-\Phi (0)=-|z|^2 +\sum_{k\geq 1}
\left (t_k z^k +\bar t_k \bar z^k \right )
\eeq
A simple calculation shows that $t_k$ are just
given by (\ref{momt}).

The basic fact of the theory of deformations of
closed analytic curves is that the (in general
complex) moments $\{ t_k, \bar t_k\}\equiv \{ t_{\pm k}\}$
supplemented by the real variable
$t_0$ form a set of local coordinates in
the ``moduli space" of smooth closed curves.
This means that under any small deformation of the domain the set
${\bf t} = \{t_0,t_{\pm k}\}$
is subject to a small change and vice
versa. For more details,
see \cite{EV,Kriunp,T}.
The differential operators
\be\label{D2}
\nabla (z)
=\p_{t_0} +\sum_{k\geq 1} \left (
\frac{z^{-k}}{k} \p_{t_k} +
\frac{\bar z^{-k}}{k}\p_{\bar t_k}\right )
\ee
span the complexified tangent space to the space of
curves.
The operator $\nabla (z)$
has a clear geometrical meaning.
To clarify it, we introduce the notion of elementary
deformation.

\begin{figure}[tb]
\epsfysize=5cm
\centerline{\epsfbox{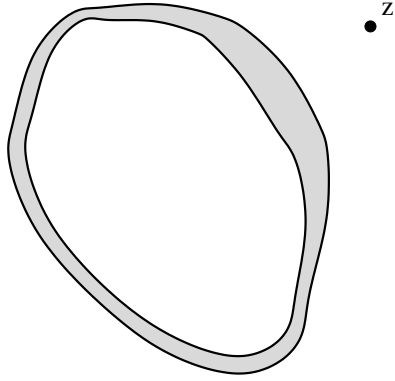}}
\caption{\sl The elementary deformation
with the base point $z$.}
\label{fi:bump}
\end{figure}

Fix a point $z\in {\sf D^c}$ and consider
a special infinitesimal
deformation of the domain such that the normal displaycement
of the boundary is proportional to the gradient of the
Green function at the boundary point
(fig.~\ref{fi:bump}):
\beq\label{small}
\delta n(\xi)
=-\frac{\epsilon}{2}\p_n G(z, \xi)
\eeq
For any sufficiently smooth initial boundary
this deformation is well-defined as $\epsilon \to 0$.
We call infinitesimal deformations from this family,
parametrized by $z\in {\sf D^c}$,
{\it elementary deformations}. The point $z$ is
refered to as the
{\it base point} of the deformation.
Note that
since $\p_n G <0$ (see the remark after the definition of the
Green function
in the Introduction), $\delta n $
for the elementary deformations is either strictly positive
or strictly negative depending of the sign of the $\epsilon$.

Let $\delta_z$ be variation of any quantity under the
elementary deformation with the base point $z$.
It is easy to see that
$\delta_z t_0 =\epsilon$,
$\delta_z t_k =
\epsilon z^{-k}/k$.
Indeed,
\be
\label{spe}
\delta_z t_k =
\frac{1}{\pi k}\oint \xi^{-k}
\delta n(\xi) |d\xi|=
-\frac{\epsilon}{2\pi k}\oint \xi^{-k}
\p_n G(z, \xi ) |d\xi|= {\epsilon \over k}\ z^{-k}
\ee
by virtue of the Dirichlet formula (\ref{Dirih}).

Let $X=X({\bf t})$ be any functional of our domain
that depends
on the harmonic moments only (in what follows
we are going to consider only
such functionals). The variation $\delta_z X$
in the leading order in $\epsilon$ is then given by
\be
\label{D4}
\delta_z X =
\sum_k \frac{\p X}{\p t_k}\, \delta_z t_k =
\epsilon \nabla (z)X
\ee
The right hand side suggests that
for functionals $X$ such that the series
$\nabla(z) X$ converges everywhere in ${\sf D^c}$ up to the
boundary, $\delta_z X$ is a harmonic function
of the base point $z$.

Note that in \cite{MWZ} we used the ``bump" deformation
and continued it harmonically to ${\sf D^c}$.
So it was the
elementary deformation (\ref{spe})
$\delta_z \propto \oint |d\xi| \p_n G(z,\xi)
\delta^{\rm bump}(\xi)$ that was
really used. The ``bump" deformation
should be understood as
a (carefully taken) limit of $\delta_z$ when the point $z$ tends
to the boundary.

\subsection{The Hadamard formula as integrability
condition}

Variation of the Green function under
small deformations of the domain is
known due to Hadamard, see eq.\,(\ref{Hadam}).
To find how the Green function changes under small
variations of the harmonic moments, we fix
three points $a,b,c \in {\bf C}\setminus {\sf D}$ and
compute $\delta_{c}G(a,b)$ by means of the
Hadamard formula (\ref{Hadam}).
Using (\ref{D4}), one can identify the result
with the action of the vector field $\nabla (c)$
on the Green function:
\be\label{Th0}
\nabla (c)G(a,b) =
-\, \frac{1}{4\pi }\oint_{\d {\sf D}}
\p_n G(a , \xi) \p_n G(b , \xi)
\p_{n} G(c , \xi) |d\xi|
\ee
Remarkably, the r.h.s. of (\ref{Th0})
is {\it symmetric} in all three arguments:
\be\label{symha}
\nabla  (a)G(b,c)
=\nabla (b)G(c,a)
=\nabla (c)G(a,b)
\ee
This is the key relation, which allows one
to represent the Dirichlet problem
as an integrable hierarchy of non-linear
differential equations \cite{MWZ}.
This relation is the integrability
condition of the hierarchy.

It follows from (\ref{symha}) (see \cite{MWZ}
for details) that there exists a function
$F=F({\bf t})$ such that
\be
\label{gf}
G(z,z') = \log\left|{1\over z} - {1\over z'}\right|
+ \ha \nabla (z)\nabla (z')F
\ee
The function $F$
is (logarithm of) the tau-function of
the integrable hierarchy. In \cite{K-K-MW-W-Z}
it was called the tau-function of the (real analytic) curves.
Existence of such a representation
of the Green function was first conjectured
by Takh\-ta\-jan. This
formula was first obtained in \cite{K-K-MW-W-Z}
(see also \cite{T} for a detailed proof
and discussion).

\subsection{Dispersionless Hirota equations for $F$}

Combining (\ref{gf}) and (\ref{Gconf}), we obtain
the relation
\beq\label{Gconf1}
\log \left |
\frac{w(z)-w(z')}{w(z)\overline{w(z')} -1} \right |^2
=\log\left|{1\over z} - {1\over z'}\right|^2
+ \nabla (z)\nabla (z')F
\eeq
which implies
an infinite hierarchy of differential equations
on the function $F$.
It is convenient to normalize
the conformal map $w(z)$ by the conditions that
$w(\infty )=\infty$ and $\p_z w(\infty )$ is real,
so that
\beq\label{confrad}
w(z)=\frac{z}{r}+O(1) \;\;\;
\mbox{as $z\to \infty$}
\eeq
where the real number
$r=\lim_{z\to\infty} {dz/ dw(z)}$
is called the
(external) conformal radius of the domain
${\sf D}$ (equivalently, it can be defined through
the Green function as
$\log r = \lim_{z\to \infty}(G(z, \infty)+\log |z|)$,
see \cite{Hille}).
Then, tending $z' \to\infty$ in
(\ref{Gconf1}),
one gets
\be\label{sec3a}
\log |w(z)|^2=\log |z|^2 - \p_{t_0}\nabla (z)F
\ee
The limit $z\to \infty$ of this equality
yields a simple formula for the conformal radius:
\be\label{sec7}
\log r^2 = \p_{t_{0}}^2 F
\ee

Let us now separate holomorphic and
antiholomorphic parts of these equations.
To do that it is convenient to introduce
holomorphic and antiholomorphic parts
of the operator $\nabla (z)$ (\ref{D2}):
\beq\label{Dhol}
D(z)=\sum_{k\geq 1}\frac{z^{-k}}{k}
\p_{t_k}\,,
\;\;\;\;\;
\bar D(\bar z)=\sum_{k\geq 1}\frac{\bar z^{-k}}{k}
\p_{\bar t_k}\,,
\eeq
Rewrite
(\ref{Gconf1}) in the form
$$
\begin{array}{ll}
&\displaystyle{
\log \left (
\frac{w(z)-w(z')}{w(z)\overline{w(z')} -1} \right )
-\log \left (\frac{1}{z}-\frac{1}{z'}\right )
-\left (\frac{1}{2}\p_{t_0} +D(z)\right )
\nabla (z' ) F } \,=
\\&\\
=&\displaystyle{
-\log \left (
\frac{\overline{w(z)}-
\overline{w(z')}}{w(z')\overline{w(z)} -1} \right )
+\log \left ( \frac{1}{\bar z}- \frac{1}{\bar z'}\right )
+\left (\frac{1}{2}\p_{t_0} +\bar D(\bar z)\right )
\nabla (z') F }
\end{array}
$$
The l.h.s. is a holomorphic function of $z$ while
the r.h.s. is antiholomorphic. Therefore, both are equal to
a $z$-independent term which can be found from the
limit $z\to \infty$.
As a result, we obtain the equation
\be\label{sec6}
\log \left (
\frac{w(z)-w(z')}{w(z)- (\overline{w(z')})^{-1} } \right )
=\log \left ( 1-\frac{z'}{z}\right )
+D(z)\nabla (z') F
\ee
which, as $z' \to \infty$, turns into the formula
for the conformal map $w(z)$:
\be\label{sec4}
\log w(z)=\log z
-\frac{1}{2}\p^{2}_{t_0}F -\p_{t_0}D(z)F
\ee
(here we used (\ref{sec7})).
Proceeding in a similar way, one can rearrange (\ref{sec6})
in order to write it separately
for holomorphic and antiholomorphic parts in $z'$:
\be\label{sec5}
\log \frac{w(z)-w(z')}{z-z'}\, =\,
- \,\frac{1}{2}\p_{t_0}^2\, F +
D(z) D(z') F
\ee
\be\label{511}
-\, \log \left (1- \frac{1}{w(z)
\overline{w(z')}}\right )=
D(z)\bar D(\bar z' )F
\ee
Writing down eqs.\,(\ref{sec5})
for the pairs of points $(a,b)$, $(b,c)$ and $(c,a)$ and
summing up the exponentials of the both sides of each equation
one arrives at the relation
\be\label{Hir1}
(a-b)e^{D(a)D(b)F}
+(b-c)e^{D(b)D(c)F}
+(c-a)e^{D(c)D(a)F} =0
\ee
which is the dispersionless Hirota equation (for the KP
part of the two-dimensional Toda lattice hierarchy)
written in the symmetric form.
This equation can be regarded
as a very degenerate case of the trisecant Fay
identity.
It encodes the algebraic relations between the second
order derivatives
of the function $F$. As $c \to \infty$, we get
these relations
in a more explicit but less symmetric form:
\be\label{Hir2}
1-e^{D(a)D(b )F}=
\frac{D(a)-D(b )}{a-b}\,\p_{t_1}F
\ee
which makes it clear that the totality of
second derivatives
$F_{ij}:=\p_{t_i}\p_{t_j}F$
are expressed through the derivatives
with one of the indices equal to unity.

More general equations of the dispersionless Toda hierarchy
obtained in a similar
way by combining eqs.\,(\ref{sec4}), (\ref{sec5}) and
(\ref{511})
include derivatives w.r.t. $t_0$
and $\bar t_k$:
\be\label{Hir3} (a-b)e^{D(a)D(b )F}
=ae^{-\p_{t_0}D(a)F}
-b e^{-\p_{t_0}D(b)F}
\ee
\be\label{Hir4}
1-e^{-D(z)\bar D(\bar z )F}=\frac{1}{z\bar z}
e^{ \p_{t_0} \nabla (z) F}
\ee
These equations allow one to express the second derivatives
$\p_{t_m}\p_{t_n}F$,
$\p_{t_m}\p_{\bar t_n}F$ with $m,n\geq 1$
through the derivatives
$\p_{t_0}\p_{t_k}F$,
$\p_{t_0}\p_{\bar t_k}F$.
In particular, the dispersionless Toda equation,
\be\label{Toda}
\p_{t_1}\p_{\bar t_1}F =e^{\p_{t_0}^{2}F}
\ee
which follows from (\ref{Hir4}) as $z \to \infty$,
expresses $\p_{t_1}\p_{\bar t_1}F$ through $\p_{t_0}^{2}F$.

For a comprehensive exposition
of Hirota equations for dispersionless
KP and Toda hierarchies we refer the reader to
\cite{C-K,T-T}.

\subsection{Integral representation of the tau-function}

Eq.\,(\ref{gf}) allows one to obtain a representation
of the tau-function as a double integral over the domain ${\sf D}$.
Set $\tilde \Phi (z):=\nabla (z)F$.
One is able to determine this function
via its variation under the elementary
deformation:
\beq\label{dual1a}
\delta_a \tilde \Phi (z) =
-2\epsilon \log \left |a^{-1}-z^{-1}\right | +
2\epsilon G(a,z)
\eeq
which is read from eq.\,(\ref{gf}) by virtue of
(\ref{D4}).
This allows one to identify $\tilde \Phi$
with the ``modified potential"
$\tilde \Phi (z)=\Phi (z)-\Phi (0) +t_0 \log |z|^2$,
where $\Phi$ is given by (\ref{dd1}).
Thus we can write
\be
\label{modpot}
\nabla (z)F =\tilde \Phi (z)=
-\frac{2}{\pi} \int_{{\sf D}} \log |z^{-1}-\zeta^{-1}| d^2 \zeta
= v_0 + 2\Re\sum_{k>0}\frac{v_k}{k}z^{-k}
\ee
The last equality is to be understood as
the Taylor expansion around infinity.
The coefficients $v_k$ are
moments of the interior domain
(the ``dual'' harmonic moments) defined as
\be
\label{vk}
v_k= \frac{1}{\pi }\int_{{\sf D}}z^{k}\,d^2 z \ \ \ (k>0)\,,
\;\;\;\;\;
v_0 =-\Phi(0)=\frac{2}{\pi}\int_{{\sf D}} \log |z|d^2 z
\ee
From (\ref{modpot}) it is clear that
\be\label{vk1}
v_k =\p_{t_k}F\,,
\;\;\;\;k\geq 0
\ee

In a similar manner, one
obtains the integral representation of the tau-function
\be\label{F}
F=-\frac{1}{\pi^2}\int_{{\sf D}} \! \int_{{\sf D}}
\log |z^{-1} -\zeta^{-1}| d^2 z d^2 \zeta
\ee
or
\be\label{F0}
F=\frac{1}{2\pi}\int_{{\sf D}}
\tilde \Phi (z) d^2 z
\ee
These formulas remain
intact in the multiply-connected case (see below).

\section{Towards the multiply-connected case and
generalized Hirota equations}

Now we are going to explain how
the above picture can be generalized
to the multiply-connected case. The details can be found in
\cite{KMZ}.

\begin{figure}[tb]
\epsfysize=7.5cm
\centerline{\epsfbox{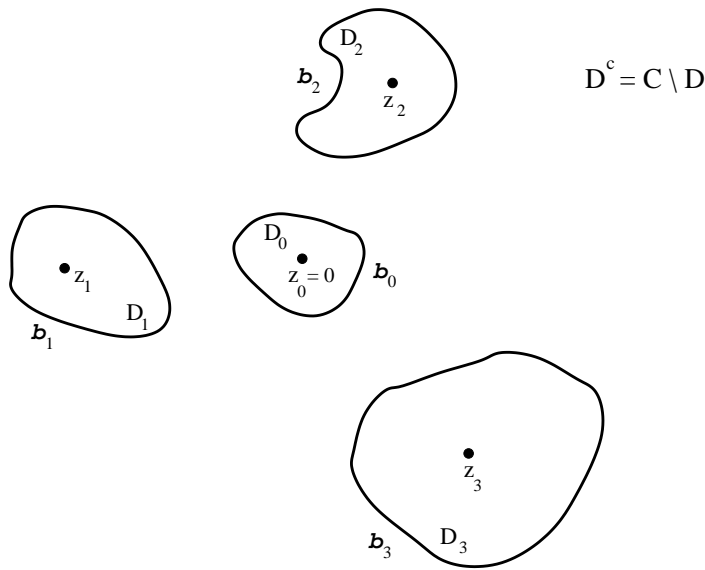}}
\caption{\sl A multiply-connected domain
${\sf D^c}={\bf C}\setminus {\sf D}$ for $g=3$.
The domain ${\sf D} = \bigcup_{\alpha=0}^3 {\sf D}_\alpha$
consists of $g+1=4$ disconnected
parts ${\sf D}_\alpha$ with the boundaries ${\sf b}_\alpha$.
To define the complete set of harmonic
moments, we also need the auxiliary points
$z_\alpha\in {\sf D}_\alpha$ which should be
always located inside the corresponding domains.}
\label{fi:multid}
\end{figure}

Let ${\sf D}_{\alpha}$, $\alpha =0, 1, \ldots , g$, be
a {\em collection} of $g+1$ non-intersecting bounded
connected domains in the complex plane
with smooth boundaries $\d {\sf D}_{\alpha}$.
Set ${\sf D}= \cup _{\alpha =0}^{g} {\sf D}_{\alpha}$,
so that the complement ${\sf D^c} = {\bf C}\setminus {\sf D}$ becomes
a multiply-connected unbounded domain in the complex plane
(see fig.~\ref{fi:multid}),
${\sf b}_{\alpha}$ being the boundary curves.

It is customary to associate with a
planar multiply-connected domain its
{\it Schottky double}, a compact Riemann surface
without boundary
endowed with an antiholomorpic involution, the boundary
of the initial domain being the set of fixed points
of the involution.
The Schottky double of the
multiply-connected domain ${\sf D^c}$ can be
thought of as two copies
of ${\sf D^c}$ (``upper" and ``lower" sheets
of the double) glued along
the boundaries
$\cup_{\alpha=0}^g {\sf b}_\alpha = \d {\sf D^c}$, with
points at infinity added
($\infty$ and $\bar \infty$).
In this set-up the holomorphic coordinate on
the upper sheet is $z$ inherited from ${\sf D^c}$,
while the holomorphic
coordinate on the other sheet is $\bar z$.
The Schottky double of
${\sf D}_c$ with two infinities
added is a compact Riemann surface of genus
$g = \#\{ {\sf D}_\alpha\}-1$.

On the double, one may choose a canonical basis
of cycles. The ${\sf b}$-cycles are just boundaries of the
holes ${\sf b}_{\alpha}$ for $\alpha =1, \ldots , g$.
Note that regarded as the oriented boundaries
of ${\sf D^c}$ (not ${\sf D}$) they have the
{\it clockwise} orientation.
The ${\sf a}_{\alpha}$-cycle connects the $\alpha$-th hole
with the 0-th one. To be more precise,
fix points $\xi_{\alpha}$ on the boundaries, then
the ${\sf a}_{\alpha}$ cycle starts from $\xi_{0}$,
goes to $\xi_{\alpha}$ on the ``upper'' (holomorphic)
sheet of the double and goes back
the same way on the ``lower'' sheet, where
the holomorphic coordinate is
$\bar z$.

\subsection{Tau-function for algebraic domains}

Comparing to the simply-connected case, nothing
is changed in posing the standard Dirichlet problem.
The definition of the Green function
and the formula (\ref{Dirih}) for the solution
of the Dirichlet problem through the Green
function are the same too.
A difference is in the nature of
harmonic functions. Any harmonic function
is the real part of an analytic function
but in the multiply-connected case
these analytic funstions are not necessarily single-valued
(only their real parts have to be single-valued).

One may still characterize the shape of a
multiply-connected domain
by harmonic moments.
However, the set of linearly independent harmonic
functions should be extended.
The complete basis of harmonic
functions in the plane with holes is described in \cite{KMZ}.

Here we shall only say a few words about the case
which requires the minimal number of
additional parameters and minimal modifications of the theory.
This is the case of {\em algebraic domains} (in the sense of
\cite{EV}), or {\em quadrature domains}
\cite{Gustafsson,Ahar-Shap},
where, roughly speaking, the space of independent
harmonic moments is
finite-dimensional. For example,
one may keep in mind the class of domains with
only finite number of non-vanishing moments.
It is this class which is directly related to multi-support
solutions of matrix models with polynomial potentials.
Boundaries of such multiply-connected domains can be
explicitly described by algebraic equations \cite{KM}.

In this case it is enough to incorporate moments
with respect to $g$ additional harmonic functions
of the form
$$
\nu_{\alpha}(z)= \log \left | 1-\frac{z_{\alpha}}{z}\right |^2 ,
\;\;\;\;\; \alpha =1,\ldots , g
$$
where $z_{\alpha}\in {\sf D}_{\alpha}$ are some marked points,
one in each hole (see fig.~\ref{fi:multid}).
Without loss of generality, it is convenient to put
$z_0 =0$. The ``periods" of these functions are:
$\oint_{{\sf b}_{\alpha}}
\d_n \nu_{\beta}(z) |dz| =4\pi \delta_{\alpha \beta}$.
The independent
parameters for algebraic domains are:
\beq\label{moments}
\begin{array}{l}
\displaystyle{
t_0 =\frac{1}{\pi}\int_{\sf D} d^2 z =
\frac{\mbox{Area}({\sf D})}{\pi}}
\\ \\
\displaystyle{
t_k =-\frac{1}{\pi k}\int_{{\sf D^c}}z^{-k} d^2 z}\,,
\;\;\;\; k\geq 1
\\ \\
\displaystyle{
\phi_{\alpha} =-\frac{1}{\pi }\int_{{\sf D^c}}
\log \left | 1-\frac{z_{\alpha}}{z}\right |^2 d^2 z}\,,
\;\;\;\; \alpha =1, \ldots , g
\end{array}
\eeq
($t_k$ are complex numbers while $t_0$ and $\phi_{\alpha}$
are real). Instead of $\phi_\alpha$ it is more convenient
to use
\be
\label{pialpha}
\Pi_{\alpha}=\phi_{\alpha}-2\,
{\rm Re}\, \sum_{k>0}t_kz_{\alpha}^k
\ee
which does not depend on the choice of $z_{\alpha}$'s.
Note that in the case of a finite number of non-vanishing
moments the sum is always well-defined.

Using the Hadamard formula, one again derives the
``exchahge relations'' (\ref{symha}) which imply the
existence of the tau-function and the
fundamental relation (\ref{gf}). They have the same
form as in the simply-connected case.
It can be shown that taking derivatives
of the tau-function with respect to the additional
variables $\Pi_{\alpha}$, one obtains the harmonic
measures of boundary components and the period matrix.

The {\it harmonic measure}
$\omega_{\alpha}(z)$
of the boundary component ${\sf b}_{\alpha}$
is the harmonic function in ${\sf D^c}$
such that it is equal to 1 on ${\sf b}_{\alpha}$
and vanishes on the other boundary curves.
From the general
formula (\ref{Dirih}) we conclude that
\be
\label{periodG}
\omega_{\alpha}(z)=
-\, \frac{1}{2\pi}\oint_{{\sf b}_{\alpha}}
\d_n G(z, \zeta )|d\zeta |,
\ \ \ \ \ \ \alpha=1,\ldots,g
\ee
Being harmonic, $\omega_{\alpha}$ can be represented
as the real part of a holomorphic function:
$$
\omega_{\alpha}(z)=W_{\alpha}(z)+\overline{W_{\alpha}(z)}
$$
where $W_{\alpha}(z)$ are holomorphic multivalued
functions in ${\sf D^c}$. The differentials
$dW_{\alpha}$
are holomorphic in ${\sf D^c}$ and purely imaginary
on all boundary contours. So they
can be extended holomorphically to the lower sheet
of the Schottky double as $-d\overline{W_{\alpha}(z)}$.
In fact this is the canonically normalized basis
of holomorphic differentials on the double.
Indeed, according to
the definitions,
$$
\oint_{{\sf a}_{\alpha}}\! dW_{\beta} =
2 {\rm Re} \int_{\xi_0}^{\xi_{\alpha}}
dW_{\beta}(z)=
\omega_{\beta}(\xi_{\alpha})\!-\!
\omega_{\beta}(\xi_0) =\delta_{\alpha \beta}
$$
Then the matrix of ${\sf b}$-periods of these differentials
reads
\be\label{Talbe}
T_{\alpha \beta}=\oint_{{\sf b}_{\alpha}}dW_{\beta} =
- \frac{i}{2}\oint_{{\sf b}_{\alpha}}\p_n \omega_{\beta} dl
=i\pi \Omega_{\alpha \beta}
\ee
The period matrix $T_{\alpha \beta}$ is purely
imaginary non-degenerate matrix with positively
definite imaginary part.
In addition to (\ref{gf}), the following relations hold:
\beq\label{u4}
\omega_{\alpha}(z)=
-\, \p_{\alpha} \nabla (z) F
\eeq
and
\beq\label{u5}
T_{\alpha \beta}=
2\pi i \,
\p_{\alpha} \p_{\beta} F
\eeq
where $\d_\alpha := \d/\d\Pi_\alpha$.

In the multiply-connected case,
the suitable analog of the conformal map
$w(z)$ (or rather of $\log w(z)$)
is the embedding of ${\sf D^c}$ into the
$g$-dimensional complex torus ${\bf Jac}$,
the Jacobi variety of the Schottky double.
This embedding is given, up to an overall
shift in ${\bf Jac}$, by the Abel map
$z \mapsto {\bf W}(z):= (W_1 (z), \ldots , W_g (z) )$
where
\beq\label{E1}
W_{\alpha}(z)=\int_{\xi_0}^{z} dW_{\alpha}
\eeq
is the holomorphic part of the
harmonic measure $\omega_{\alpha}$.
By virtue of (\ref{u4}), the Abel map
is represented through the second order derivatives
of the function $F$:
\beq\label{E2}
W_{\alpha}(z)-W_{\alpha}(\infty)=
\int_{\infty}^{z} dW_{\alpha}=
-\p_{\alpha}D(z) F
\eeq
\beq\label{E3}
2\, {\rm Re}\, W_{\alpha}(\infty)=
\omega_{\alpha}(\infty)=-\p_{t_0}\p_{\alpha}F
\eeq
The last formula immediately follows from (\ref{u4}).

\subsection{The Green function and the
generalized Hirota equations.}

The Green function of the Dirichlet
boundary problem in the multiply-connected case,
can be written in terms of the prime form (see \cite{Fay} for the definition
and properties)
on the Schottky double (cf. (\ref{Gconf})):
\be
\label{prime}
G(z,\zeta) = \log\left|E(z,\zeta)\over E(z, \bar\zeta)\right|
\ee
Here by $\bar\zeta$ we mean the (holomorphic) coordinate of
the ``mirror" point on the Schottky double, i.e.
the ``mirror" of $\zeta$ under the
antiholomorphic involution.
Using (\ref{prime}) together with (\ref{gf}), (\ref{u4}) and (\ref{u5})
one can obtain the following
representations of the prime form in terms of the tau-function
\be\label{E10-12}
E(z, \zeta )=(z^{-1}-\zeta^{-1})
e^{-\frac{1}{2}(D(z)-D(\zeta ))^2 F}
\\
iE(z, \bar \zeta )=
e^{-\frac{1}{2}(\p_{t_0} +D(z)+ \bar D(\bar \zeta ))^2 F}
\\
iE (z, \bar z )=
e^{-\frac{1}{2}\nabla^2 (z) F}
\ee
generalizing (\ref{sec5}), (\ref{511}) in the simply-connected case.

This allows us to write the generalized Hirota equations for
$F$ in the
multiply-connected case.
They follow from the Fay identities \cite{Fay} and
(\ref{E10-12}). In analogy to the simply-conected case,
any second order derivative of the function
$F$ w.r.t. $t_k$ (and $\bar t_k$), $F_{ik}$,
is expressed through the
derivatives
$\{ F_{\alpha\beta}\}$ where
$\alpha,\beta=0,\ldots,g$ together
with $\{ F_{\alpha t_i}\}$ and their complex conjugated.
To be more
precise, one can consider all second derivatives as functions of
$\{ F_{\alpha\beta}, F_{\alpha k} \}$ modulo
certain relations on
the latter discussed in \cite{KMZ}; sometimes on this
``small phase space" more extra
constraints arise, which can be written in the form similar to the
Hirota or WDVV equations \cite{MMMBr}.

For the detailed discussion of the
generalized Hirota relations
the reader is addressed to \cite{KMZ}.
Here we just give the simplest example of such relations,
an analog of the dispesrionless Toda equation (\ref{Toda})
for the
tau-function. It reads
\beq\label{Todagen}
\p_{t_1} \p_{\bar t_1} F
=\frac{\theta (\bomega (\infty ) \!+\!\bZ )
\theta (\bomega (\infty )\!-\! \bZ )}{\theta^2 (\bZ )}
\, e^{\p_{t_0}^{2}F}\,
+\!\! \sum_{\alpha , \beta =1}^{g}
\! (\log \theta (\bZ ))_{, \, \alpha \beta}
\, (\p_{\alpha} \p_{t_1}F)
(\p_{\beta} \p_{\bar t_1}F)
\eeq
Here $\theta$ is the Riemann theta-function with the
period matrix $T_{\alpha \beta}$ and
$$
(\log \theta (\bZ ))_{, \, \alpha \beta} :=
\p^2 \log \theta (\bZ )/\p Z_{\alpha} \p Z_{\beta}
$$
The equation holds for any
vector-valued parameter ${\bf Z}\in {\bf Jac}$.
It is important to note
that the theta-functions are expressed through
the second order derivatives of $F$, so
(\ref{Todagen}) is indeed a partial differential
equation for $F$. For example,
$$
\theta \left (\bomega (\infty )\right )=
\sum_{n_{\alpha}\in \bZ}\exp \left (
 -2\pi^2 \sum_{\alpha \beta}n_{\alpha}n_{\beta}
\p_{\alpha \beta}^{2} F -
2\pi i \sum_{\alpha}n_{\alpha}
\p_{\alpha}\p_{t_0}  F \right )
$$


\section{Conclusion}

In these notes we have reviewed the integrable structure of the Dirichlet
boundary problem. We have presented the simplest known to us proof of the
Hadamard variational formula and derivation of the dispersionless Hirota
equations for the simply-connected case.

We have also demonstrated how this approach can be generalized to the case
of multiply-connected domains. The main ingredients remain intact, but the
conformal map to the reference domain should be substituted by the Abel map
into Jacobian of the Schottky double of the multiply-connected domain. Then
one can write the generalization of the Hirota equations using the Fay
identities, a particular case of which leads to generalization of the
dispersionless Toda equation.

Here we have only briefly commented on the properties of the
quasiclassical tau-function of the multiply-connected solution. A detailed
discussion of this issue and many related problems, including conformal maps
in the multiply-connected case, duality transformations on the Schottky
double, relation to the multi-support solutions of the matrix models etc,
can be found in \cite{KMZ}.

\section*{Acknowledgments}

We are indebted to I.Kri\-che\-ver and P.Wieg\-mann
for collaboration on different stages of this work and to
V.Ka\-za\-kov, A.Le\-vin,
M.Mi\-ne\-ev-\-Wein\-stein, S.Na\-tan\-zon and
L.Takh\-ta\-jan and illuminating discussions.
The work was partialy supported by RFBR under
the grants 00-02-16477,
by INTAS under the grant 99-0590
and by the Program of support of scientific schools
under the grants 1578.2003.2 (A.M.), 1999.2003.2 (A.Z.).
The work of A.Z. was also partially supported by
the LDRD project 20020006ER ``Unstable Fluid/Fluid Interfaces''.
We are indebted to P.~van~Moerbeke for
encouraging us to write this
contribution and to F.~Lambert for the warm hospitality on Elba
during the conference.

\end{document}